\documentclass[prb,preprint,amsmath,amssymb]{revtex4}

\usepackage{graphicx}
\usepackage{bm}

\newcommand{\lang}{\left\langle}
\newcommand{\rang}{\right\rangle}

\begin{document}

\title{Free energy surfaces from nonequilibrium processes without work measurement}

\author{Artur B. Adib}
\email{artur@brown.edu}
\affiliation{
Department of Physics and Department of Chemistry, Box 1843, Brown University, Providence, Rhode Island 02912, USA
}

\date{\today}

\begin{abstract}
Recent developments in statistical mechanics have allowed the estimation of equilibrium free energies from the statistics of work measurements during processes that drive the system out of equilibrium. Here a different class of processes is considered, wherein the system is prepared and released from a nonequilibrium state, and no external work is involved during its observation. For such ``clamp-and-release'' processes, a simple strategy for the estimation of equilibrium free energies is offered. The method is illustrated with numerical simulations, and analyzed in the context of tethered single-molecule experiments.
\end{abstract}

\maketitle


\section{Introduction}

Since the seminal work of Jarzynski in 1997,\cite{jarzynski97-prl} considerable attention has been directed towards the study of exact and directly measurable equalities relating equilibrium free energies (or ``potentials of mean force'') and nonequilibrium measurements (see e.g. Refs.~\onlinecite{crooks00,hummer01-pnas,hendrix01,shirts03,dellago05,liphardt02,collin05}). In broad terms, these equalities allow one to compute potentials of mean force along a reaction coordinate from the statistics of work measurements in a class of nonequilibrium processes, wherein the reaction coordinate is dynamically driven between configurations of interest by some external agent. Apart from its intrinsic value as a formal generalization of the second law of thermodynamics, the possibility of estimating equilibrium free energy differences in this manner bears practical consequences in both computational and experimental settings; in the context of computer simulations, for example, the Jarzynski equality\cite{jarzynski97-prl} (JE) and the Crooks fluctuation theorem\cite{crooks00}  (CFT) allow the use of ``embarrassingly parallel'' implementations,\cite{hendrix01} lowest-variance estimators,\cite{shirts03} and unusually large time-steps,\cite{dellago05} to name a few, while in the experimental context they offer a convenient framework to extract free energy information from work measurements in processes that are most naturally carried out in a non-quasistatic fashion, such as ``pulling'' experiments with single biological molecules.\cite{liphardt02,collin05}

In the present work, it will be shown that an analogous procedure yielding free energy differences from nonequilibrium measurements exists in a different class of processes, where the reaction coordinate is no longer dynamically driven between the configurations of interest, and hence no work or force measurement is involved. In order to introduce this process and its associated equality, it is convenient to specialize the analysis to the ``tethered'' single-molecule experiment shown in Figure~\ref{illustration}, which is of easy visualization and direct relevance to a number of ongoing experiments with biomolecules (see below). In this kind of experiment, the end-to-end distance $q$ of the molecule (henceforth referred to as the {\em reaction coordinate}) is controlled by some external device, which is capable of performing two simple tasks: a ``clamp'' operation, whereby the reaction coordinate is placed and held fixed at a position of interest for as long as one desires, and a ``release'' operation, in which the external agent stops acting on the molecule, allowing the reaction coordinate to take on any value. It is also assumed that one has the ability to observe the position of the reaction coordinate at an arbitrary time $\tau$ after it is released by the external agent. The combined use of these operations will be referred to as a {\em clamp-and-release} process.

Consider a series of clamp-and-release processes performed in the following sequence: (a) The reaction coordinate is clamped at a value of interest $q_A$ until the remaining degrees of freedom of the molecule equilibrate with the environment. (b) The coordinate is then released at some arbitrary time $t=0$, and its position $q$ recorded at a later time $t=\tau$. Iteration of (a)-(b) a number of times allows one to construct a histogram of $q$ that approaches a probability distribution, $p_\tau(q|q_A)$, in the limit of an infinite number of realizations. In this notation, the function $p_\tau(q'|q)$ specifies the probability that the reaction coordinate takes on the value $q'$, $\tau$ units of time after it being released from the clamped and equilibrated position $q$. Once this histogram is constructed to the desired accuracy, a second batch of experiments is performed, where steps (a)-(b) are carried out with another position of interest $q_B$ in place of $q_A$. Similarly, this defines another probability distribution, $p_\tau(q|q_B)$. The central result of this work is that these two probability distributions (or, approximately, their respective histograms) are directly related to the free energy difference $\Delta f = f_B - f_A$ between the two configurations of interest, $q_B$ and $q_A$ respectively, as
\begin{equation} \label{the-eq}
  \frac{p_\tau(q_B|q_A)}{p_\tau(q_A|q_B)} = e^{-\beta \Delta f}, \quad (\forall \, \tau>0)
\end{equation}
where $\beta = 1/k T$ is the temperature parameter involving the Boltzmann constant $k$ and the temperature $T$ at which the molecule was equilibrated. As this result holds {\em at all times} $\tau > 0$, one can also write it as
\begin{equation} \label{the-eq-sum}
  \frac{1}{M} \sum_{i=0}^M \frac{p_{\tau_i}(q_B|q_A)}{p_{\tau_i}(q_A|q_B)} = e^{-\beta \Delta f},
\end{equation}
where the histograms of the reaction coordinate are computed at $M$ different times $\tau_i > 0$. This property enhances the efficiency of the method by taking advantage of the time-series of $q$, rather than a single snapshot thereof, and will be used in the numerical applications below.

Several remarks can be made regarding Eq.~(\ref{the-eq}) before embarking on its derivation and numerical verification. As a quick consistency check, observe that Eq.~(\ref{the-eq}) reduces to the correct equilibrium result when $\tau \to \infty$, as in this limit the motion of $q$ is uncorrelated with its originally clamped position, i.e. 
\begin{equation*}
  \frac{ p_\tau(q_B|q_A) } {p_\tau(q_A|q_B)} \to \frac{p_{eq}(q_B)}{p_{eq}(q_A)}, \quad (\tau \to \infty)
\end{equation*}
where $p_{eq}(q)$ is the equilibrium probability distribution of $q$; since the equilibrium distribution of the reaction coordinate defines the potential of mean force (cf. Eq.~(\ref{pmf})), this validates Eq.~(\ref{the-eq}) in the long-time limit. Like the JE and the CFT, the clamp-and-release processes that underly this equality take place in nonequilibrium conditions: as already observed, only after some relaxation time will the molecule explore its allowed conformations according to the equilibrium distribution. However, unlike the JE and the CFT which externally drive the reaction coordinate between the states $q_A$ and $q_B$, Eq.~(\ref{the-eq}) does not immediately guarantee that the molecule will overcome (after some time $\tau$) the possible free energy barriers higher than $kT$ that might exist between the configurations $q_A$ and $q_B$. Though likely to be milder in single-molecule experiments, this problem is particularly severe in computer simulations, where such barriers essentially prevent the molecule from going between the configurations $q_A$ and $q_B$ in a feasible time-scale. Nonetheless, both in the computational and in the experimental context, this {\em rare event} difficulty is easily overcome by performing additional clamp-and-release experiments at intermediate points between $q_A$ and $q_B$; for example, if one adds an intermediate point $q_C$ between $q_A$ and $q_B$ so that the odds of the transitions $q_A \leftrightarrow q_C$ and $q_C \leftrightarrow q_B$ are greater than those of $q_A \leftrightarrow q_B$, the desired free energy difference $\Delta f = f_B - f_A$ can be computed by simply adding the intermediate free energy differences, i.e. $\Delta f = (f_C - f_A) + (f_B - f_C)$, where the free energy differences in parentheses are computed from the corresponding intermediate clamp-and-release experiments. In addition to providing a more detailed landscape of the free energy surface between $q_A$ and $q_B$, the information contained in the intermediate histograms lends itself to a simple strategy allowing one to make educated choices of additional intermediate points as the experiment is performed (see below).

\section{Proof}

Although a full proof of Eq.~(\ref{the-eq}) from first principles will be offered shortly, it is interesting to note that this result follows almost immediately from Onsager's statement of {\em microscopic reversibility} (cf. paragraph after Eq.~(4.2) of Ref.~\onlinecite{onsager31b}, or Ref.~\onlinecite{mazur84} for a modern proof), which in the present context and notation reads
\begin{equation} \label{onsager}
  p_{eq}(q,t_1;q',t_2) = p_{eq}(q',t_1;q,t_2),
\end{equation}
where $p_{eq}(q,t_1;q';t_2)$ is the joint probability -- in equilibrium -- that the reaction coordinate takes the value $q$ at time $t_1$, {\em and} the value $q'$ at time $t_2>t_1$. (Note that this statement is strictly distinct from Onsager's regression hypothesis, also put forward in the same paper above, which relates the decay of equilibrium correlations with nonequilibrium relaxation processes, and only applies to small external disturbances\cite{jarzynski-onOnsager}). From elementary probability theory, one can decompose joint probabilities in terms of the associated marginal and conditional probabilities, e.g.
\begin{equation*}
  p_{eq}(q,t_1;q',t_2) = p_{eq}(q) \, p_{eq}(q',t_2|q,t_1),
\end{equation*}
respectively, where $p_{eq}(q',t_2|q,t_1)$ is the conditional probability that the reaction coordinate will take on the value $q'$ at $t_2$ given that it had the value $q$ at $t_1$. Upon substituting this decomposition in Eq.~(\ref{onsager}), we arrive at Eq.~(\ref{the-eq}), with $\tau \equiv t_2 - t_1$ and the identification $p_\tau(q'|q) = p_{eq}(q',\tau+t_1|q,t_1)$.

As hinted by the above derivation based on Onsager's reversibility principle, Eq.~(\ref{the-eq}) ought to depend upon the property of time-reversal symmetry of the microscopic equations of motion. There are several deterministic and stochastic models that embody this property, and Eq.~(\ref{the-eq}) should be insensitive to the specific model chosen. In fact, Eq.~(\ref{onsager}) follows directly from the time-reversal property of time-correlation functions. For example, in Ref.~\onlinecite{adib05-chem} a time-reversible Hamiltonian model was used to derive a symmetry relation in chemical kinetics that is essentially Eq.~(\ref{the-eq}); in particular, that derivation reveals the validity of this equation even when the molecule is coupled to a non-Markovian heat bath. The derivation offered below is somewhat simpler, as it assumes a Markovian heat bath and discrete time (see e.g. Ref.~\onlinecite{crooks00} for a similar approach), but the generality of Eq.~(\ref{the-eq}) should be born in mind. This derivation also makes direct contact with Monte Carlo simulations, which will be used later in the paper to illustrate the validity and use of Eq.~(\ref{the-eq}).

In accordance with the above discussion, let us begin the derivation of Eq.~(\ref{the-eq}) by defining the clamp-and-release probability as
\begin{multline} \label{ptau-def}
  p_\tau(q_B|q_A) \equiv \int \! dx_0 \cdots dx_\tau \, \frac{ \rho_{eq}(x_0) \delta(q(x_0) - q_A) } {p_{eq}(q_A)}  \\
    \times P(x_1|x_0) \cdots P(x_t|x_{t-1}) \delta(q(x_\tau) - q_B),
\end{multline}
where $x_t$ is the phase space vector containing all the degrees of freedom of the molecule at time $t=0,1,\ldots,\tau$, $P(x'|x)$ is the dynamic transition rule of the Markov chain dictating the probability of going from state $x$ to state $x'$,\cite{norris97} and $q(x_t)$ is the function that yields the value of the reaction coordinate for a given configuration $x_t$. Note that the initial statistical state is correctly normalized by the factor $p_{eq}(q_A)$, where
\begin{equation*}
  p_{eq}(q') = \int \! dx \, \rho_{eq}(x) \, \delta(q(x) - q'),
\end{equation*}
and $\rho_{eq}(x)$ is the unconstrained equilibrium distribution of the molecule. In this stochastic model, the property of microscopic reversibility is often expressed dynamically by the ``detailed balance'' condition, namely\cite{norris97,crooks99}
\begin{equation} \label{detbalance}
  \rho_{eq}(x) P(x'|x) = \rho_{eq}(\overline{x}') P(\overline{x}|\overline{x}'), \quad (\forall \, x,x')
\end{equation}
where $\overline{x}$ indicates the time-reversed counterpart of $x$ in which all components of $x$ that are odd under time-reversal (such as momenta) have opposite sign.\cite{crooks99} This ensures that the probability of observing the trajectory $x_0,x_1,\ldots,x_\tau$ is identical to the probability of observing its time-reversed counterpart $\overline{x}_\tau,\overline{x}_{\tau-1},\ldots,\overline{x}_0$, i.e.
\begin{equation*}
  \rho_{eq}(x_0) P(x_1|x_0) \cdots P(x_\tau|x_{\tau-1}) = \rho_{eq}(\overline{x}_\tau) P(\overline{x}_{\tau-1}|\overline{x}_\tau) \cdots P(\overline{x}_0|\overline{x}_1),
\end{equation*}
as can be straightforwardly verified by successive application of Eq.~(\ref{detbalance}) in the left hand side of the above equation. Using this result in Eq.~(\ref{ptau-def}) and rearranging terms, we have
\begin{multline} 
  p_\tau(q_B|q_A) = \frac{p_{eq}(q_B)}{p_{eq}(q_A)} \int \! dx_\tau \cdots dx_0 \, \frac{ \rho_{eq}(\overline{x}_\tau) \delta(q(x_\tau) - q_B) } {p_{eq}(q_B)}  \\
    \times P(\overline{x}_{\tau-1}|\overline{x}_\tau) \cdots P(\overline{x}_0|\overline{x}_1) \delta(q(x_0) - q_A).
\end{multline}
Upon changing all integration variables to their time-reversed counterparts (this comes with a Jacobian of unity) and using the fact that the end-to-end distance is invariant under time-reversal, i.e. $q(x)=q(\overline{x})$, one can recognize the resulting integral as $p_\tau(q_A|q_B)$. The remaining ratio of equilibrium probabilities is related to the potentials of mean force through the usual definition
\begin{equation} \label{pmf}
  e^{-\beta f(q)} \equiv p_{eq}(q),
\end{equation}
which finally yields Eq.~(\ref{the-eq}).

\section{Numerical results}

To illustrate the validity and use of Eqs.~(\ref{the-eq})-(\ref{the-eq-sum}), the free energy surface of a one-dimensional model comprised of a particle experiencing a barrier of height $\sim 20 kT$ was reconstructed (see Ref.~\onlinecite{hummer01-pnas} for a similar approach). The specific profile of the potential is shown in Figure~\ref{results}, along with the free energy surface reconstructed via Eq.~(\ref{the-eq-sum}) with a total of $20$ points between the two minima of the potential (as only free energy {\em differences} are available, one point is lost to set the ``zero'' of the vertical axis). Note that the free energy surface coincides with the potential energy, in accordance with the dimensionality of the problem. In these Monte Carlo (MC) experiments, $100$ trajectories of $100$ MC time-steps each were generated with the standard Metropolis algorithm\cite{frenkel02} from each clamped position, and $M=10$ observations equally spaced in time of the coordinate $q$ were used to construct the histograms required by Eq.~(\ref{the-eq-sum}).

Though the agreement observed in Figure~\ref{results} illustrates the validity of Eqs.~(\ref{the-eq}) and (\ref{the-eq-sum}), it offers little insight into the advantages and drawbacks of the method. In particular, it is important to understand the aforementioned rare event issue in more detail. To this end, snapshots of the distribution of the reaction coordinate after being released from two nearby positions (namely, $q=-0.6$ and $q=-0.5$) are shown in Figure~\ref{snapshots}. For these histograms, 10,000 trajectories were used so as to obtain smoother curves, although the general profile of the distributions is essentially unchanged with only 100 trajectories. The first direct observation is that both distributions are moving towards the nearest minimum at $q=-1$ (as indicated by the horizontal arrows), in agreement with the expected behavior due to the positive slope of the potential between $q=-1$ and $q=0$. This asymmetry of visits to the left and to the right of the original clamped position is a manifestation of the rare event issue: $p_\tau(-0.5|-0.6)$ is much smaller than $p_\tau(-0.6|-0.5)$ (cf. circles in Figure~\ref{snapshots}). However, this drawback can be turned into a diagnosis tool: suppose, for example, that one had measured $p_\tau(q|-0.6)$, and that one were sweeping the clamped positions of the reaction coordinate from left to right. The form of the distribution $p_\tau(q|-0.6)$ can be used to make an educated choice for the next point to the right of $q=-0.6$, e.g. by establishing a threshold value of the probability and choosing the value of $q$ that corresponds to this threshold. This ensures that one will be spending most of the time and effort in regions dominated by steep energy barriers, while alleviating the effort in regions where the distributions move about in a more symmetric fashion.

\section{Applicability in single-molecule experiments}

\begin{figure}
\begin{center}
\includegraphics[width=350pt]{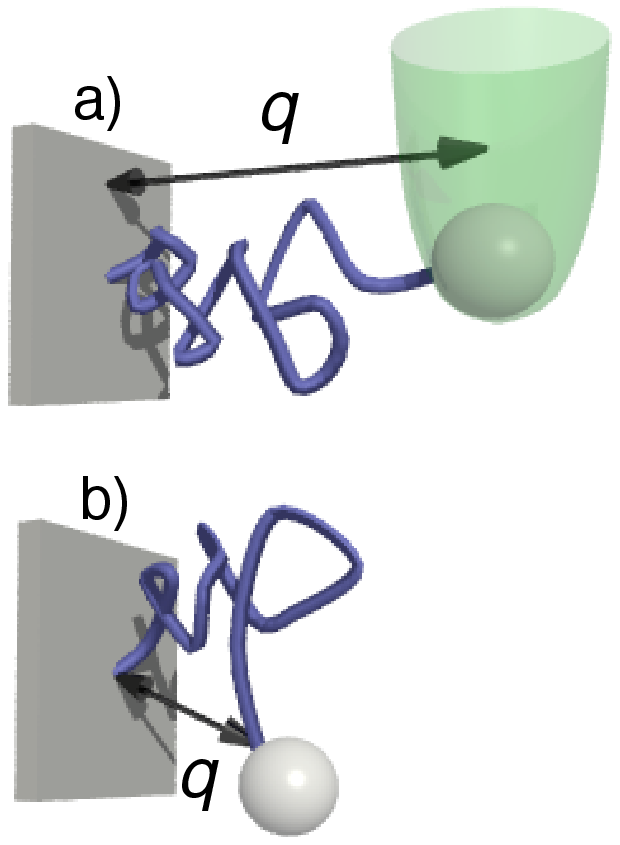} 
1
\label{illustration}
\end{center}
\end{figure}

The above results suggest that the clamp-and-release method might be useful in single-molecule experiments, which are becoming a standard research tool in various laboratories worldwide (see e.g. Ref.~\onlinecite{bustamante00-review} for a review). In these experiments, one is able to isolate, manipulate, and to a certain extent ``observe'' an individual biological molecule (e.g. DNAs, RNAs, and proteins) at the micron-to-nanometer level, through a setup much like the one shown in Figure~\ref{illustration}. Although the application of Eqs.~(\ref{the-eq}) and (\ref{the-eq-sum}) is in principle straightforward in this context, there are some discrepancies between the idealized clamp-and-release process introduced in connection with these equalities, and the actual processes and measurements one is presently able to perform in such experiments. Here only a few of these issues will be discussed, namely the stiffness of the trapping potential, the ability to observe the reaction coordinate $q$ in a time-resolved manner, and the effects of the bead on the dynamics and statistics of the molecule. These items will be discussed separately below.

In tethered single-molecule experiments, one routinely manipulates the position of polystyrene beads attached to the molecules of interest by means of laser ``tweezers,'' which effectively induce a harmonic confining potential on the bead.\cite{bustamante00-review} Associated with this confining potential is the existence of thermal fluctuations about its minimum, thus in principle violating the premise of the clamp operation. However, a back-of-the-envelope estimate based on the equipartition theorem\cite{bustamante00-review} suggests that such spatial fluctuations are orders of magnitude smaller than the nanometer scale one is often interested, namely $\lang \Delta x^2 \rang \sim 10^{-11}-10^{-17}$ m, and hence negligible in the present context (assuming room temperature and the stiffness range $10^{-10}-10^{-4}$ N m$^{-1}$ encountered in typical optical traps\cite{bustamante00-review}). Note that more careful studies in the context of the JE (where a general result explicitly accounting for the external potential exists\cite{hummer01-pnas}) have led to similar conclusions.\cite{schurr03,hummer05}

As the present method requires time-resolved observations of the reaction coordinate, the ability to track the position of the tethered bead is an essential ingredient for its successful implementation. This can be achieved through the rapidly evolving single-particle tracking (SPT) techniques, which have been widely used in the context of tethered single-molecule experiments for over a decade (see e.g. Refs.~\onlinecite{schafer91,finzi95,zocchi01,oddershede02,pouget04,blumberg05}). The use of SPT combined with optical traps has already been adopted for quantitative studies of the motion of proteins in the outer membrane of {\em E. coli} bacteria,\cite{oddershede02} where the $\lambda$-receptor protein was tethered to a polystyrene bead, which was in turn manipulated with a laser trap and observed with SPT techniques. A more recent study has reported the use of evanescent-wave methods\cite{zocchi01} to obtain a full three-dimensional characterization of the bead position with excellent spatial (nm) and temporal ($30-350$ Hz) resolution.\cite{blumberg05} Together with laser trap techniques, this setup seems to offer a natural setting for the use of the clamp-and-release methodology, which can induce barrier-crossing events (i.e. new conformations) otherwise not observed in a feasible time-scale without the clamp stage.

Finally, a crucial question is whether the presence of the tethered bead affects the properties of the standalone molecule one is ultimately interested in. For example, it is clear that the dynamics of the molecule (and hence that of $q$) is appreciably affected by the presence of the relatively heavy bead, which also experiences a viscous drag as it travels through the surrounding fluid.\cite{qian99} Moreover, even when it is held in place by an external field, the presence of the bead might change the allowed conformations of the molecule of interest by simple volume exclusion.\cite{segall05} Instead of trying to characterize all such effects separately, it is best to view the results of tethered single-molecule experiments as probing either dynamical or equilibrium properties of the bead-molecule complex. Thus, although the dynamics of the complex differs from that of the standalone molecule in a non-trivial fashion,\cite{qian99} the {\em equilibrium} properties (such as free energy differences) are essentially affected by volume-exclusion only. This has been the subject of a recent theoretical investigation,\cite{segall05} which concluded that -- in the particular context of Figure~\ref{illustration}, where one end of the molecule is attached to a glass slab -- the size of the tethered bead does influence the equilibrium properties of the molecule. However, these effects are expected to be minimized when both ends of the molecule are attached to beads, as in this case the motion of the tethered bead is no longer constrained to one side of a plane only, but a more careful study in this case does not seem to be available. More importantly, in this particular two-bead context, a technique that further minimizes volume-exclusion effects is the ``molecular handles'' approach of C. Bustamante and collaborators,\cite{liphardt02,collin05} in which the molecule of interest is connected to the beads through parts of DNA or DNA-RNA hybrids. In this case, a systematic study has shown that the effects of the handles and of the bead surfaces on the properties of the molecule of interest are negligible, at least under typical experimental conditions, and provided the handles are long enough.\cite{bustamante05-private} This technique also seems to be an important ingredient for the success of the clamp-and-release method.

\section{Conclusions}

In summary, a nonequilibrium method for the estimation of equilibrium free energies based on a ``clamp-and-release'' methodology has been put forward, and analyzed both in computational and experimental contexts. Underpinning the method is a simple equality based on the property of time-reversal symmetry (Eq.~(\ref{the-eq})), which relates equilibrium free energies to repeated observations of the reaction coordinate as the molecule relaxes from a clamped nonequilibrium state. This equality and the associated clamp-and-release process differ fundamentally from existing nonequilibrium methods based on the Jarzynski equality or the Crooks fluctuation theorem, as no work or force measurement is involved. Experimentally, this might be particularly useful when the external device controlling the reaction coordinate is not directly capable of performing force measurements, as in the case of the new generation of nanotweezers based on carbon nanotubes.\cite{bustamante00-review,kim99} Computationally, the method offers a conceptually simple algorithm for the estimation of free energy differences that is particularly suited for problems involving the crossing of high free energy barriers along a known reaction coordinate (cf. strategy based on intermediate points above), which complements existing equilibrium and nonequilibrium methods.\cite{frenkel02} It remains to be seen how the efficiency of the present method ranks against such methodologies.

\acknowledgments

The author is indebted to Prof. Jimmie Doll for his guidance and advice. Discussions and/or correspondences with Prof. Carlos Bustamante, Dr. Gavin Crooks, and Dr. Christopher Jarzynski were also fruitful. This research was supported by the US Department of Energy under grant DE-FG02-03ER46074.

\newpage

{\bf Figure captions}

\vspace{1cm}

FIG 1: Illustration of a ``tethered'' single-molecule experiment (not to scale) relevant to the clamp-and-release method. In both cases (a) and (b), one end of the linear molecule is attached to a nonreactive bead while the other end is attached to a fixed surface, which can be either a glass slab or another bead held by a micropipette (not shown). (a) Initially, the reaction coordinate $q$ is clamped by means of an external device, here represented by an effective potential energy surface surrounding the bead. (b) The molecule is then released by the external device, thus allowing $q$ to take on any value.

\vspace{1cm}

FIG 2: Potential energy (solid line) used in the Monte Carlo simulations described in the text, and the ``free energy surface'' (circles) reconstructed via Eq.~(\ref{the-eq-sum}), with $M=10$, and 100 trajectories for each point. The bin size of the histograms was $\Delta q = 0.1$.

\vspace{1cm}

FIG 3: Histograms of the reaction coordinate at two different times, $t=30$ time-steps (upper box), and $t=100$ time-steps (lower box), from an MC simulation with 10,000 trajectories. For each time, two distributions are shown: one that was originally clamped at $q=-0.6$ (solid curves), and another at $q=-0.5$ (dashed curves). The vertical lines specify the original clamped positions of the reaction coordinate at $q=-0.6$ (solid lines) and $q=-0.5$ (dashed lines), while the circles indicate the intersection of a given curve with the original clamped position of the other, yielding the probabilities $p_\tau(-0.6|-0.5)$ and $p_\tau(-0.5|-0.6)$. The horizontal arrows represent the displacement of the peak of the distributions from their original positions at $t=0$.

\newpage

\newpage

\begin{figure}
\begin{center}
\includegraphics[width=350pt]{results.eps} 
2
\label{results}
\end{center}
\end{figure}

\newpage

\begin{figure}
\begin{center}
\includegraphics[width=350pt]{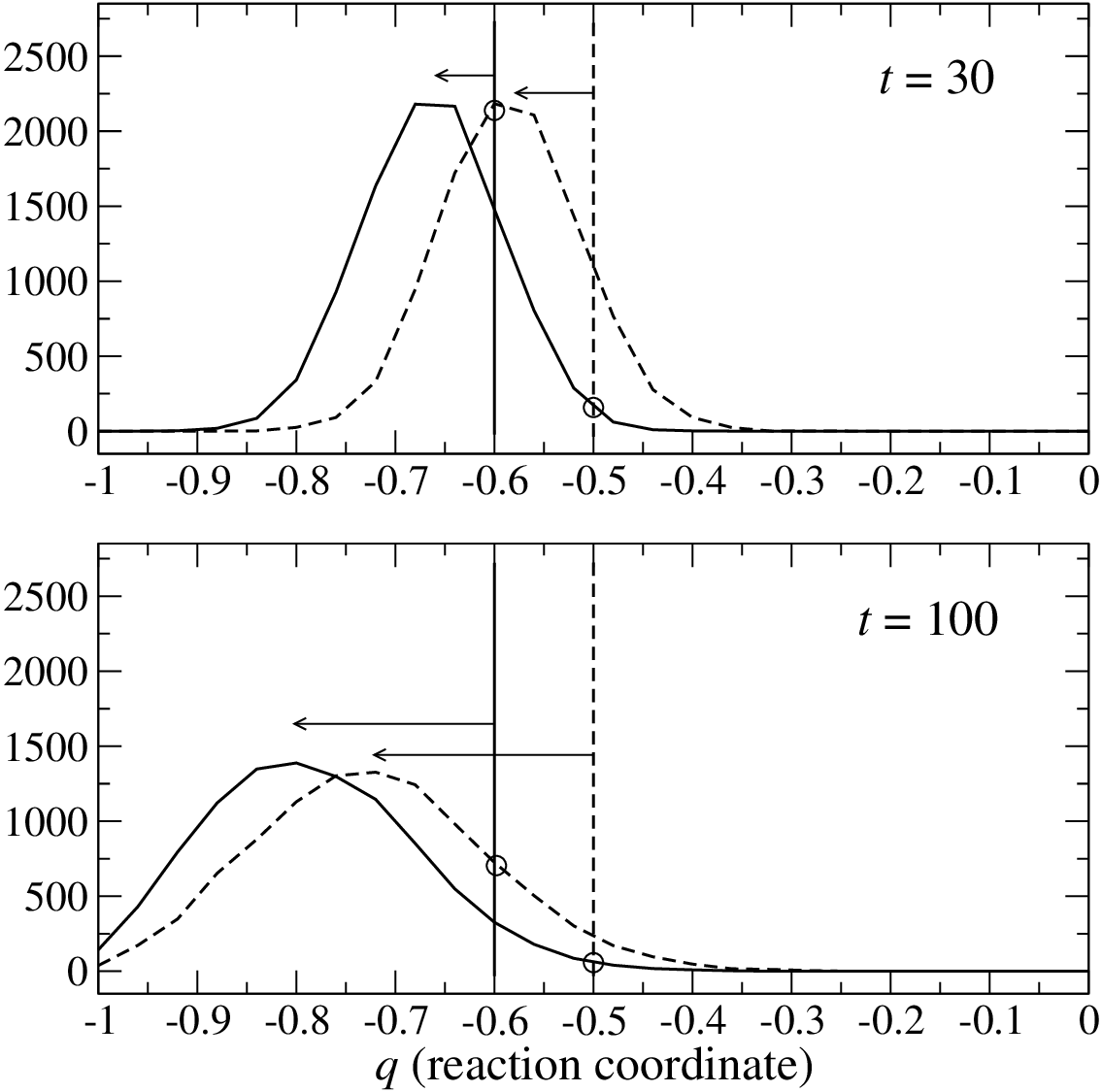} 
3
\label{snapshots}
\end{center}
\end{figure}


\begin{thebibliography}{28}
\bibitem[{jar()}]{jarzynski-onOnsager}
\bibinfo{note}{Private conversations with Christopher Jarzynski were very
  instructive in elucidating the contents of Onsager's original paper.}
\end{thebibliography}
\end{document}